\documentclass[aps,twocolumn,showpacs,preprintnumbers,amsmath,amssymb,superscriptaddress,floatfix,nofootinbib]{revtex4}

\usepackage{graphicx}
\usepackage{hyperref}
\usepackage{amsmath}
\allowdisplaybreaks[1]
\usepackage{amsfonts}
\usepackage{amssymb}
\usepackage{multirow}
\usepackage{makecell}
\usepackage[dvipsnames]{xcolor}
\usepackage{booktabs}
\usepackage{tabularx}
\usepackage{makecell}
\usepackage{ulem}
\usepackage{float}

\usepackage[caption=false]{subfig}

\newcommand*{\Rom}[1]{\uppercase\expandafter{\romannumeral #1\relax}}

\begin{document}

\title{$P_{cc}^N$ states in a unitarized coupled-channel approach}

\author{Chao-Wei Shen} \email{c.shen@fz-juelich.de}
\affiliation{Institute for Advanced Simulation, Institut f\"ur Kernphysik and
J\"ulich Center for Hadron Physics, \\ Forschungszentrum J\"ulich, D-52425 J\"ulich, Germany}

\author{Yong-hui Lin} \email{yonghui@hiskp.uni-bonn.de}
\affiliation{Helmholtz-Institut f\"ur Strahlen- und Kernphysik and Bethe Center for
Theoretical Physics,\\ Universit\"at Bonn, D-53115 Bonn, Germany}

\author{Ulf-G. Mei{\ss}ner} \email{meissner@hiskp.uni-bonn.de}
\affiliation{Helmholtz-Institut f\"ur Strahlen- und Kernphysik and Bethe Center for
Theoretical Physics,\\ Universit\"at Bonn, D-53115 Bonn, Germany}
\affiliation{Institute for Advanced Simulation, Institut f\"ur Kernphysik and J\"ulich
Center for Hadron Physics, \\ Forschungszentrum J\"ulich, D-52425 J\"ulich, Germany}
\affiliation{Tbilisi State University, 0186 Tbilisi, Georgia}

\date{\today}

\begin{abstract}
Starting from an effective Lagrangian with heavy quark spin symmetry embedded, the coupled-channel
dynamics of the doubly charmed systems $D^{(*)} \Sigma_c^{(*)}$ is investigated. The potential
underlying our investigation includes $t$-channel pseudoscalar and vector meson exchanges.
A series of $S$-wave bound states with isospin $I=1/2$ is found by applying the first iterated
solution of the $N/D$ method: one state with binding energy $23$~\,MeV in the $5/2^-$ $D^*\Sigma_c^*$
channel, three states with binding energy $26$, $30$ and $7$\,MeV (relative to the thresholds from
low to high, respectively) in the $3/2^-$ $D\Sigma_c^*$-$D^*\Sigma_c$-$D^*\Sigma_c^*$ system and
three states with binding energy $32$, $8$ and $16$\,MeV in the $1/2^-$
$D\Sigma_c$-$D^*\Sigma_c$-$D^*\Sigma_c^*$ system. Those $P_{cc}^N$ states serve as the open-charm
partners of the hidden charm pentaquarks $P_{\psi}^N$ observed by the LHCb Collaboration.
\end{abstract}

\maketitle

\section{Introduction}

Searches of exotic hadrons, whose valence quark composition is beyond the conventional picture
where mesons and baryons are composed of a pair of quark-antiquark ($q\bar{q}$) and three quarks
($qqq$), respectively, have become an important project for most of the collider facilities
especially after the experimental observations of tetraquark and pentaquark
candidates~\cite{Belle:2003nnu,BaBar:2005hhc,Belle:2011aa,BESIII:2013ris,BESIII:2013ouc,LHCb:2015yax,COMPASS:2015kdx}.
In 2015, the LHCb Collaboration announced the first evidence of two hidden-charm pentaquark-like
states $P_{\psi}^N(4380)$ and $P_{\psi}^N(4450)$ in the $J/\psi p$ invariant mass spectrum
measured from the decay process $\Lambda_b^0\to J/\psi K^- p$~\cite{LHCb:2015yax}.
In 2019, the mass spectrum of $P_{\psi}^N$ pentaquarks was updated to three states, that is,
$P_{\psi}^N(4312)$, $P_{\psi}^N(4440)$ and $P_{\psi}^N(4457)$ by the LHCb Collaboration~\cite{LHCb:2019kea}.
In 2020, a new hidden-charm pentaquark state with strangeness, namely $P_{\psi s}^{\Lambda}(4459)$, was
observed in the $J/\psi \Lambda$ invariant mass distribution from the $\Xi_b^-\to J/\psi \Lambda K^-$
decay~\cite{LHCb:2020jpq}.
And very recently, another hidden-charm pentaquark with strangeness, $P_{\psi s}^{\Lambda}(4338)$, was
announced by the LHCb Collaboration~\cite{LHCb:2022}.
It is observed in the invariant mass spectrum of $J/\psi \Lambda$ in the decay $B^-\to
J/\psi \Lambda \bar{p}$. In the last decade, the LHCb Collaboration has found many surprises
in exotic hadron spectroscopy. One can expect that the richness of the exotic spectrum will
continue to increase in the foreseeable future.

The theoretical investigations on the exotic spectroscopy date back to the birth of the
quark model in 1964. The existence of pentaquark states was first pointed out by Gell-Mann in his
famous paper on the quark model~\cite{Gell-Mann:1964ewy}.
From the point of view of modern physics, neither the multiquark states that made up of
more than three valence quarks such as tentraquarks ($qq\bar{q}\bar{q}$) and pentaquarks
($qqqq\bar{q}$), nor the hybrid states that have both valence quarks and gluons or the glueballs
that are composed of pure valence gluons are forbidden by Quantum Chromodynamics (QCD),
which is the fundamental theory of the strong interactions.
Before the first experimental evidence of $P_{\psi}^N$, these pentaquark states have been
predicted by the theoretical work based on the phenomenological coupled-channel approach in
2010~\cite{Wu:2010jy}. In that work, one $\bar{D}\Sigma_c$ and one $\bar{D}^*\Sigma_c$ bound state
were found around 4\,GeV, which can be related to the observed $P_{\psi}^N(4312)$ and $P_{\psi}^N(4440)$
or $P_{\psi}^N(4457)$ states, respectively.
In particular, the newly reported $P_{\psi s}^\Lambda(4459)$ and $P_{\psi s}^\Lambda(4338)$ states are
also compatible with the predicted $\bar{D}^*\Xi_c$ and $\bar{D}\Xi_c$ bound states, respectively.
Such impressive consistence between the experimental observations and the theoretical predictions on
the pentaquark spectrum around 4\,GeV is a strong indication for the molecular nature of those
pentaquark-like states. The hadronic molecule picture has become are much discussed approach
to explain the nature of the exotic candidates, as seen by the many  theoretical studies of exotic
hadron spectroscopy during the last decades, see the recent reviews in
Refs.~\cite{Chen:2016qju,Guo:2017jvc,Liu:2019zoy,Brambilla:2019esw,Meng:2022ozq}.

In Ref.~\cite{Dong:2020hxe}, the authers claim that the near-threshold structures exist generally
in two heavy hadron systems as long as the interaction between them is attractive.
Hundreds of hadronic molecules are proposed in the
heavy-heavy~\cite{Chen:2021htr,Dong:2021bvy,Chen:2021kad} and heavy-antiheavy
sectors~\cite{Dong:2021juy}.
And very recently, several five-flavored bound states are predicted in the $B^{(*)}\Xi_c^{(\prime)}$
system~\cite{Shen:2022rpn}.
Among these exotic baryons, the doubly-charmed pentaquarks are straightforward extensions of the
$P_{\psi}^N$ states, whose quark content can be written as $ccqq\bar{q}$~($q=u,d$).
In this work, we explore the mass spectrum of the doubly-charmed pentaquark-like states
around 4\,GeV. Some earlier investigations on the this system are given in
Refs.~\cite{Chen:2021htr,Dong:2021bvy,Chen:2021kad}.
Ref.~\cite{Chen:2021htr} constructed the contact, one-pion-exchange, and two-pion-exchange
potentials for the coupled-channel $D^{(*)} \Sigma_c^{(*)}$ system within the framework of a
chiral effective field theory and found the $S$-wave bound states by solving the nonrelativistic
Schr\"{o}dinger equation.
Ref.~\cite{Chen:2021kad} updated the configuration in Ref.~\cite{Chen:2021htr} by introducing
the $S$-$D$ mixing effect within the one-boson-exchange (OBE) model assisted with heavy quark
spin symmetry (HQSS) and Ref.~\cite{Dong:2021bvy} only studied the single channel case.
All those three works give a similar mass spectrum for the doubly-charmed pentaquark candidates,
that is, one $1/2^-$ $D\Sigma_c$, one $3/2^-$ $D\Sigma_c^*$, two $D^*\Sigma_c$ with spin-parity
$1/2^-$ and $3/2^-$, and three $D^*\Sigma_c^*$ bound states with spin-parity $1/2^-$, $3/2^-$ and $5/2^-$.
All those doubly-charmed bound states have isospin $I=1/2$ and the binding energies var from several
MeV to tens of MeV.
%


In the present work, both pseudoscalar ($\pi$, $\eta$) and vector ($\rho$, $\omega$) meson exchanges
are considered by means of an effective Lagrangians that is constrained by the HQSS together with
the chiral symmetry for the pseudoscalar meson part and the hidden local symmetry for the vector
meson part. The bound states and resonances are found as poles of the coupled-channel scattering
amplitudes given by a unitarized Bethe-Salpeter equation (BSE) in the on-shell factorization approach.
Further, the first iterated solution of the $N/D$ method is employed to avoid the unphysical left-hand-cut
problem in the on-shell factorized BSE~\cite{Gulmez:2016scm,Du:2018gyn}.

This work is organized as follows. In Sec.~\ref{Sec:theory}, we present the theoretical framework
of our calculation.
The numerical results for the $D \Sigma_c$, $D \Sigma_c^*$, $D^* \Sigma_c$ and $D^* \Sigma_c^*$
coupled-channel dynamics and relevant discussions are presented in Sec.~\ref{Sec:result}.
Finally, a brief summary is given in Sec.~\ref{Sec:summary}. Some technicalities are
relegated to the Appendices.
%


\section{Formalism} \label{Sec:theory}

Before going to the details of our theoretical calculations, we briefly count the number of channels
in the coupled-channel $D^{(*)} \Sigma_c^{(*)}$ system included in our work.
It should be noticed that we only consider $S$-wave scattering with isospin $I=1/2$ throughout
the present work.
The quantum numbers of the various $D^{(*)} \Sigma_c^{(*)}$ channels are listed in Tab.~\ref{Tab:channels}.
It shows that we have three channels ($D\Sigma_c$, $D^*\Sigma_c$ and $D^*\Sigma_c^*$) for
spin-parity $J^P=1/2^-$, three channels ($D\Sigma_c^*$, $D^*\Sigma_c$ and $D^*\Sigma_c^*$) for
$J^P=3/2^-$, and one channel ($D^*\Sigma_c^*$) for $J^P=5/2^-$.

\begin{table}[htbp]
	\centering
	\renewcommand\arraystretch{1.3}
	\caption{$S$-wave channels in the $D \Sigma_c$-$D \Sigma_c^*$-$D^* \Sigma_c$-$D^* \Sigma_c^*$
          coupled-channel system. The quantum numbers are given in the spectroscopic form ${}^{2S+1}L_{J}$.
          \label{Tab:channels}}
	\begin{tabular}{p{1.7cm}<{\centering}p{1.9cm}<{\centering}p{1.7cm}<{\centering}p{1.7cm}<{\centering}}
		\hline
		\hline
		Channel & $J^P=1/2^-$ & $3/2^-$ & $5/2^-$ \\
		\hline
		$D\Sigma_c$  & ${}^2S_{1/2}$  & -	&	- \\
		$D\Sigma_c^*$&	-	&	${}^4S_{3/2}$	&	-	\\
		$D^*\Sigma_c$  & ${}^2S_{1/2}$  & ${}^4S_{3/2}$	&	- \\
		{$D^*\Sigma_c^*$}  & ${}^2S_{1/2}$  & ${}^4S_{3/2}$	&	${}^6S_{5/2}$ \\
		\hline
		\hline
	\end{tabular}
\end{table}

\subsection{Effective Lagrangians and the on-shell factorization approach of the Bethe-Salpeter equation}\label{subsec_potential}

Chiral perturbation theory (ChPT) developed in the 1980s~\cite{Weinberg:1978kz,Gasser:1983yg,Gasser:1984gg}
has achieved great success in describing low-energy experiments of the strong interaction,
especially the $\pi\pi$ and $\pi N$ scattering~\cite{Bernard:2006gx}. A variety of ChPT
variants were proposed to solve various specific strong-interaction systems.
Among them, the heavy baryon and heavy meson chiral perturbation theory are designed to
describe the interactions between two hadrons containing one or more heavy
quarks~\cite{Wise:1992hn,Yan:1992gz,Cheng:1992xi,Casalbuoni:1996pg}, see Ref.~\cite{Meng:2022ozq}
for a recent review.
Similar to the ChPT language, the interactions between two heavy hadrons are constructed from
light pseudoscalar exchange, the Goldstone bosons from the spontaneous breaking of chiral symmetry.
The interactions between heavy hadrons and light vector mesons are built by using the hidden local
symmetry approach~\cite{Bando:1984ej,Bando:1987br,Meissner:1987ge}.
All the relevant effective Lagrangians used here are given as~\cite{Dong:2021juy,Liu:2011xc}
\begin{align}
    {\cal L}_1 =& i g \langle H_b^{(Q)} {\cal A}_{ba}^\mu \gamma_\mu \gamma_5 \bar{H}_a^{(Q)} \rangle, \nonumber \\
    {\cal L}_2 =& -i\beta \langle H_b^{(Q)} v^\mu \rho_{\mu,ba} \bar{H}_a^{(Q)} \rangle \nonumber \\
&+ i \lambda \langle H_b^{(Q)} \sigma^{\mu\nu} (\partial_\mu \rho_{\nu} - \partial_\nu \rho_{\mu})_{ba} \bar{H}_a^{(Q)} \rangle, \nonumber \\
    {\cal L}_S =& \frac32 g_1(iv_\kappa) \epsilon^{\mu\nu\lambda\kappa} \bar{S}_{\mu,ab}^{(Q)} {\cal A}_{\nu,bc} S_{\lambda,ca}^{(Q)} - i \beta_S \bar{S}_{\mu,ab}^{(Q)} v^\alpha \rho_{\alpha,bc} \nonumber \\
&\times S_{ca}^{(Q)\mu} + \lambda_S \bar{S}_{\mu,ab}^{(Q)} (\partial^\mu \rho^{\nu} - \partial^\nu \rho^{\mu})_{bc} S_{\nu,ca}^{(Q)}, 
\label{Eq:Lag}
\end{align}
where the Lorentz indices are given by the greek letters $\mu,\nu,\cdots$, the SU(3) flavor indices
are denoted by latin symbols $a, b, \cdots$ and $\langle \cdots \rangle$ is the Dirac trace.
The summation over repeated indices is implicit.
The explicit formulae for all vertices in our $t$-channel potentials are then obtained by
expanding the above Lagrangians.
The expansion of all the underlying Lagrangian and the definitions of the involved field operators
are displayed in App.~\ref{App:formalism}.

The scattering amplitude $T$ is unitarized through the Bethe-Salpeter equation, namely
\begin{equation}\label{eq:BSE}
	T=V+VGT,
\end{equation}
where $V$ is the scattering kernel that is expressed in terms of the $t$-channel one-boson-exchange
transitions between two channels and $G$ is the two-meson loop function.
In the coupled-channel case, $T$ is a $n\times n$ matrix ($n$ denotes the number of the
coupled channels) and $G$ becomes a $n$-dimension diagonal matrix with all the elastic
loop functions $g_i$ as its elements ($i$ is the channel index).
Working with the on-shell factorization approach, the integral equation~\eqref{eq:BSE} is
reduced to an algebraic equation and one can solve the amplitude $T$ with
\begin{equation}\label{eq:on-shell}
	T=(1-VG)^{-1}V.
\end{equation}
Unitarity and analyticity of $T$ are guaranteed in the on-shell factorization prescription
in most cases~\cite{Oller:1997ng,Oller:1998hw,Oller:1998zr}.
And it has been applied commonly into various hadron systems to study the low-energy strong
interaction dynamics in them, see
e.g.~\cite{Oller:1997ti,Oller:2000fj,Lutz:2001yb,Garcia-Recio:2002yxy,Jido:2003cb,Hyodo:2007jq,Guo:2012yt,Albaladejo:2016lbb,Guo:2017jvc,Meng:2022ozq}
(and references therein).
It should also be mentioned that in some cases where the unphysical left-hand cuts (usually come from
the partial-wave projection of $V$) are not far away from the energy regions of interest,
the unitarity and analyticity of $T$ become problematic due to the existence of such
unphysical cuts, see e.g. Refs.~\cite{Du:2017ttu,Du:2018gyn} for more details.

For the two-meson Green's function $g_i$, we adopt dimensional regularization to arrive at
\begin{align}
g_i =& i \int \frac{d^4q}{(2\pi)^4} \frac1{(p-q)^2-M_B^{i2}+i\varepsilon}
\frac1{q^2-M_P^{i2}+i\varepsilon} \nonumber \\
=& \frac{1}{16\pi^2} \Big\{ a(\mu) + \ln\frac{M_B^{i2}}{\mu^2} + \frac{M_P^{i2}-M_B^{i2}+s}{2s}
\ln\frac{M_P^{i2}}{M_B^{i2}} \nonumber \\
&+ \frac{\bar{q}_i}{\sqrt s} \big[ \ln(s-M_B^{i2}+M_P^{i2}+2\bar{q}_i\sqrt s) \nonumber \\
&+ \ln(s+M_B^{i2}-M_P^{i2}+2\bar{q}_i\sqrt s) \nonumber \\
&- \ln(-s-M_B^{i2}+M_P^{i2}+2\bar{q}_i\sqrt s) \nonumber \\
&- \ln(-s+M_B^{i2}-M_P^{i2}+2\bar{q}_i\sqrt s) \big] \Big\}, \label{eq:G_RS1}
\end{align}
where $s=p^2$ and $$\bar{q}_i=\sqrt{\frac{(s-(M_B^{i}+M_P^{i})^2)(s-(M_B^{i}-M_P^{i})^2)}{4 s}}.$$
$M_B^{i}, M_P^{i}$ denote the baryon and meson masses in the channel $i$, and
$a(\mu)$ is the scale-dependent subtract constant.
Note that ${\rm Im}(\bar{q})\ge0$ indicates that Eq.~\eqref{eq:G_RS1} gives the loop function
in the physical sheet, denoted as $g_i^{\Rom{1}}$.
The loop function in the unphysical sheet, denoted as $g_i^{\Rom{2}}$, is then expressed as
\begin{equation}
	g_i^{\Rom{2}}(s)=g_i^{\Rom{1}}+2 i \rho_i^{}(s),
\end{equation}
with $\rho_i(s)=\bar{q}_i/(8\pi\sqrt{s})$.
Another strategy commonly adopted to calculate the loop function $G$ is to introduce a
phenomenological form factor, such as the Gaussian regulator, in the integral, that is,
\begin{align}
g_i =& \int \frac{q^2 dq}{4\pi^2} \frac{\omega_B^i+\omega_P^i}{\omega_B^i \omega_P^i}
\frac{e^{-2q^2/\Lambda^2}}{s-(\omega_B^i+\omega_P^i)^2+i\varepsilon},
\label{eq:G2}
\end{align}
where $\omega_B^i$ and $\omega_P^i$ are the on-shell energies for the baryon and meson in $i$th channel,
respectively.
In this work, we take the convention of Eq.~\eqref{eq:G_RS1}, where $a(\mu)$ is estimated by
matching $g_i$ of Eq.~\eqref{eq:G_RS1} to the one of Eq.~\eqref{eq:G2} at $i$th threshold with the
empirical values of cutoff $\Lambda$, i.e., around $1$~GeV.

\subsection{First iterated solution of the $N/D$ method}\label{subsec_ND}

As mentioned above, the existence of the left-hand cuts (LHC) could invalidate the on-shell
factorization formula of Eq.~\eqref{eq:on-shell} in some cases.
Unfortunately, this indeed happens in the $D^{(*)}\Sigma_c^{(*)}$ systems of interest.
In this subsection, we briefly introduce the $N/D$ method that can be used to treat those
unphysical LHC properly~\cite{Chew:1960iv,Bjorken:1960zz,Oller:1998zr,Gulmez:2016scm,Du:2018gyn}.

In the $N/D$ method, the unitarized scattering amplitude $T$ is constructed through the
dispersion relations and it has the general form
\begin{equation}
	T(s)=D(s)^{-1}N(s),
\end{equation}
where the numerator $N(s)$ and denominator $D(s)$ contain the analytic information of the left-
and right-hand cuts, respectively.
The general expressions of $N(s)$ and $D(s)$ for the $S$-wave are given by
\begin{align}\label{eq:n/d}
	D(s)&=\sum_{m=0}^{n-1}a_m s^m + \frac{(s-s_0)^n}{\pi}\int_{s_{\rm thr}}^\infty d s^\prime \frac{\rho(s^\prime)N(s^\prime)}{(s^\prime-s)(s^\prime-s_0)^n},\notag\\
	N(s)&=\sum_{m=0}^{n-1}b_m s^m +\frac{(s-s_0)^n}{\pi}\int_{-\infty}^{s_{\rm left}} d s^\prime \frac{{\rm Im} T(s^\prime)D(s^\prime)}{(s^\prime-s)(s^\prime-s_0)^n},
\end{align}
where the polynomials $\sum_m a_m s^m$ and $\sum_m b_m s^m$ denote the subtraction terms with
$a_m$ and $b_m$ the corresponding subtraction constants. $s_0$ is the subtraction point and
$n$ is the number of subtractions that is required to ensure the convergence of the dispersion integrals.
Note that the so-called Castiliejo-Dalitz-Dyson (CDD) poles~\cite{Castillejo:1955ed} are dropped here.
The difficulty caused by the unphysical LHC will be overcome if one solves exactly the $N/D$ integral
equations Eq.~\eqref{eq:n/d}.
An approximative strategy called the first-iterated solution of the $N/D$ method that was
proposed in Refs.~\cite{Gulmez:2016scm,Du:2018gyn} is utilized in our work.
It states that we approximate the numerator $N(s)$ as the tree-level potential $V(s)$ and then
the denominator $D(s)$ can be expressed as
\begin{align}\label{eq:dfun1}
	D_{ij}(s)&=\gamma_{0ij}+\gamma_{1ij}(s-s_{\rm thr}^j)+\frac12\gamma_{2ij}(s-s_{\rm thr}^j)^2\notag\\
	&+\frac{(s-s_{\rm thr}^j)s^2}{\pi}\int_{s^j_{\rm thr}}^\infty d s^\prime\frac{V_{ij}(s^\prime)
        \rho_j(s^\prime)}{(s^\prime-s_{\rm thr}^j)(s^\prime-s)s^{\prime 2}},
\end{align}
where $i$ and $j$ are the channel indices.
The subtraction point $s_0$ is set to be $s_{\rm thr}^j$, the $j$th threshold, and $n=3$.
Three subtraction constants $\gamma_{0ij}$, $\gamma_{1ij}$ and $\gamma_{2ij}$ are determined by
matching $D_{ij}(s)$ of Eq.~\eqref{eq:dfun1} and $\delta_{ij}-V_{ij}(s)G_{j}$ around the threshold
$s^j_{\rm thr}$ for each $i$ and $j$, specifically.
We fit $D_{ij}(s)$ of Eq.~\eqref{eq:dfun1} to $\delta_{ij}-V_{ij}(s)G_{j}$ in the small
energy region from the threshold $s^j_{\rm thr}$ to $100$\,MeV above it.
It is worth mentioning that $G_{j}$ in the matching procedure is the loop function in
the physical sheet. Consequently, the scattering amplitude $T$ of the physical sheet is
calculated with the $D$ function constructed in Eq.~\eqref{eq:dfun1}.
Further, $T$ in the unphysical sheets are defined as~\cite{Du:2018gyn},
\begin{equation}\label{eq:T_RS2}
	T^{\Rom{2}}(s)=\frac{1}{[T^{\Rom{1}}(s)]^{-1}-2i\rho},
\end{equation}
where $\rho$ denotes the diagonal matrix ${\rm diag}\{N_i\rho_i(s)\}$ with $N_i=0$ and $1$
representing the physical and unphysical sheets for the $i$th channel.
The uncertainty stemming from the ambiguity of the choice of the matching energy region will be discussed
when we present our numerical results.
%

\section{Results and discussions} \label{Sec:result}

\begin{figure}[bthp]
	\centerline{\includegraphics*[width=0.47\textwidth,angle=0]{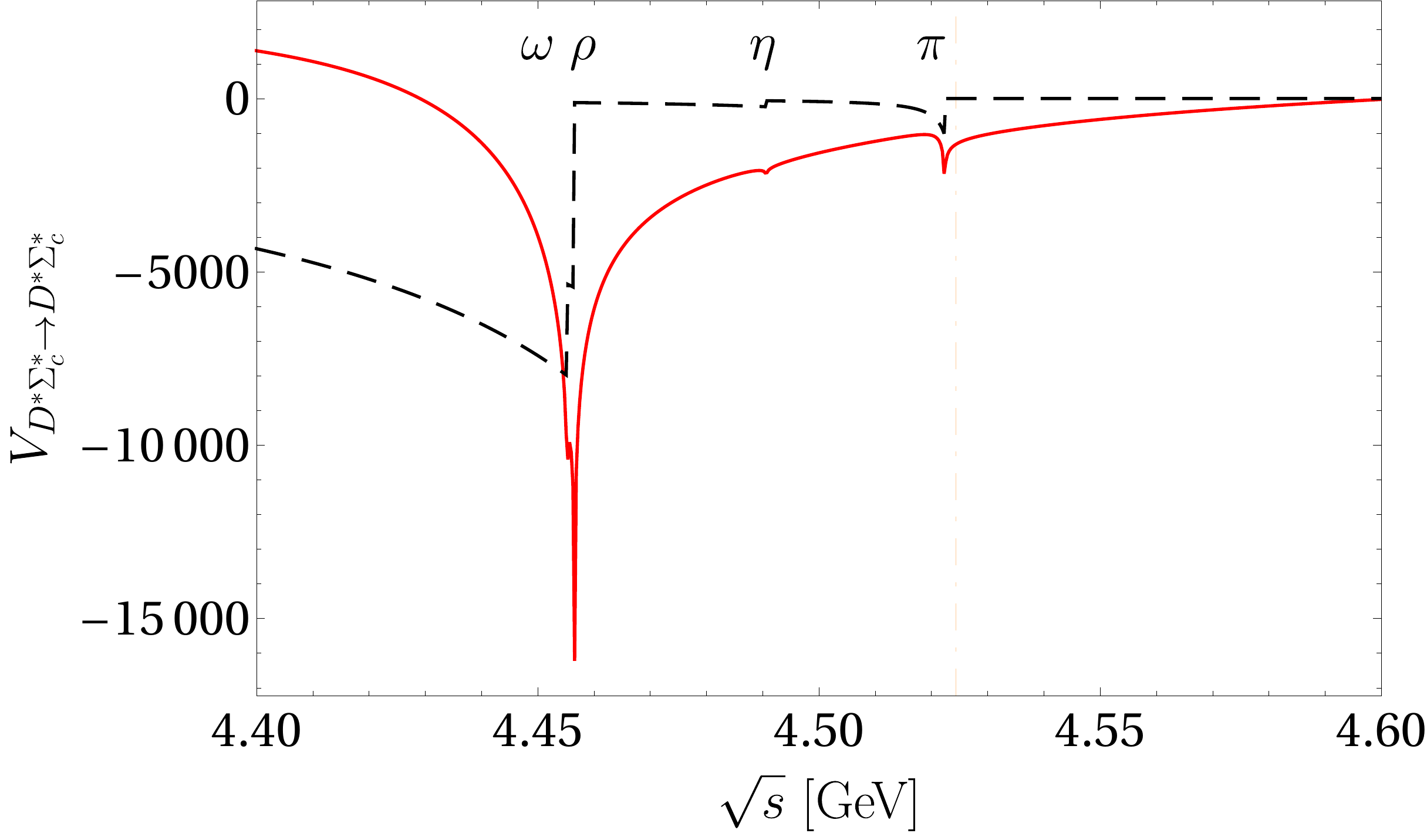}}
	\caption{$S$-wave potential for $D^*\Sigma_c^*\to D^*\Sigma_c^*$ with $(I,J)=(1/2,5/2)$
          (real part: solid curve, imaginary part: dashed curve). The corresponding threshold
          is represented by the orange dash-dotted line. The left-hand cuts from left to
          right are generated by the $\omega$, $\rho$, $\eta$ and $\pi$ exchange, respectively.
	}
	\label{fig:v55}
\end{figure}

First, we focus on the case of $J^P=5/2^-$ and explain the necessity of treating the LHC in
the partial-wave projected potentials by using the $N/D$ method.
The $S$-wave potential for the $5/2^-$-$D^*\Sigma_c^*$ system is given in terms of
the $t$-channel $\rho$-, $\omega$-, $\pi$- and $\eta$-exchange diagrams.
The numerical potential $V_{D^*\Sigma_c^*\to D^*\Sigma_c^*}$ is presented in Fig.~\ref{fig:v55}.
One can clearly see that the $S$-wave projection for each boson exchange produces one LHC
located below the $D^*\Sigma_c^*$ threshold.
Among them, the LHC corresponding to the $\pi$ exchange is quite close to the threshold and
that from the vector meson exchange is stronger.

\begin{widetext}

	\begin{figure}[H]
		\centerline{\includegraphics*[width=0.99\textwidth,angle=0]{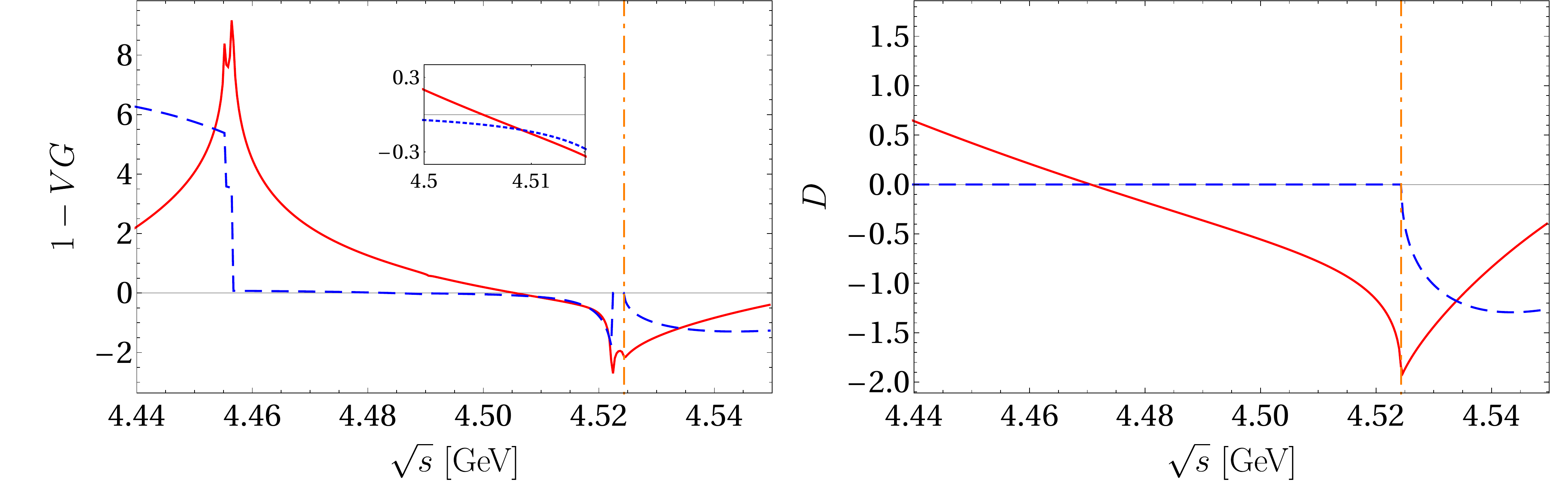}}
		\caption{$(1-VG)$ (left panel) and $D$ (right panel) in the physical sheet for
                  the $S$-wave $D^*\Sigma_c^*$ system with $(I,J)=(1/2,5/2)$ (real part: solid curve,
                  imaginary part: dashed curve). The cutoff is $\Lambda=0.7$\,GeV. The corresponding
                  threshold is represented by the orange dash-dotted line.}
		\label{fig:d55}
	\end{figure}

\end{widetext}

The $(1-VG)$ term in Eq.~\eqref{eq:on-shell} and $D$ in Eq.~\eqref{eq:dfun1} of the physical Riemann 
sheet ${\rm RS}(+)$ with the cutoff $\Lambda=0.7$\,GeV for the
$S$-wave $5/2^-$-$D^*\Sigma_c^*$ system are shown in the left and right panel of
Fig.~\ref{fig:d55}, respectively. The pole position of scattering amplitude $T$ in the single
channel case is represented by the zeros of $(1-VG)$ in the on-shell factorization BSE approach
or equivalently the zeros of the $D$ function in the $N/D$ method.
%

It can be seen from Fig.~\ref{fig:d55} that there no pole appears in the function $1-VG$ and
one pole located below the threshold at the real axis can be found in the $D$ function which is
related to a $D^*\Sigma_c^*$ bound state with binding energy $E_B=53$\,MeV.
However, all the previous works imply the existence of an $S$-wave $D^*\Sigma_c^*$ bound state
with $J^P=5/2^-$~\cite{Chen:2021htr,Dong:2021bvy}.
To be more specific, Ref.~\cite{Chen:2021htr} claimed an $S$-wave $D^*\Sigma_c^*$ state
with binding energy being around $20$\,MeV in the non-relativistic one- and two-pion exchange
potentials. Ref.~\cite{Dong:2021bvy} found a bound state from the non-relativistic $\rho$- and
$\omega$-exchange interactions with the binding energy varying from $2$ to $40$\,MeV.
The $S$-wave $D^*\Sigma_c^*$ state with $E_B=53$\,MeV obtained from the $N/D$ method is consistent
with the previous non-relativistic investigations.
%

\begin{widetext}

	\begin{figure}[t!]
		\centerline{\includegraphics*[width=0.99\textwidth,angle=0]{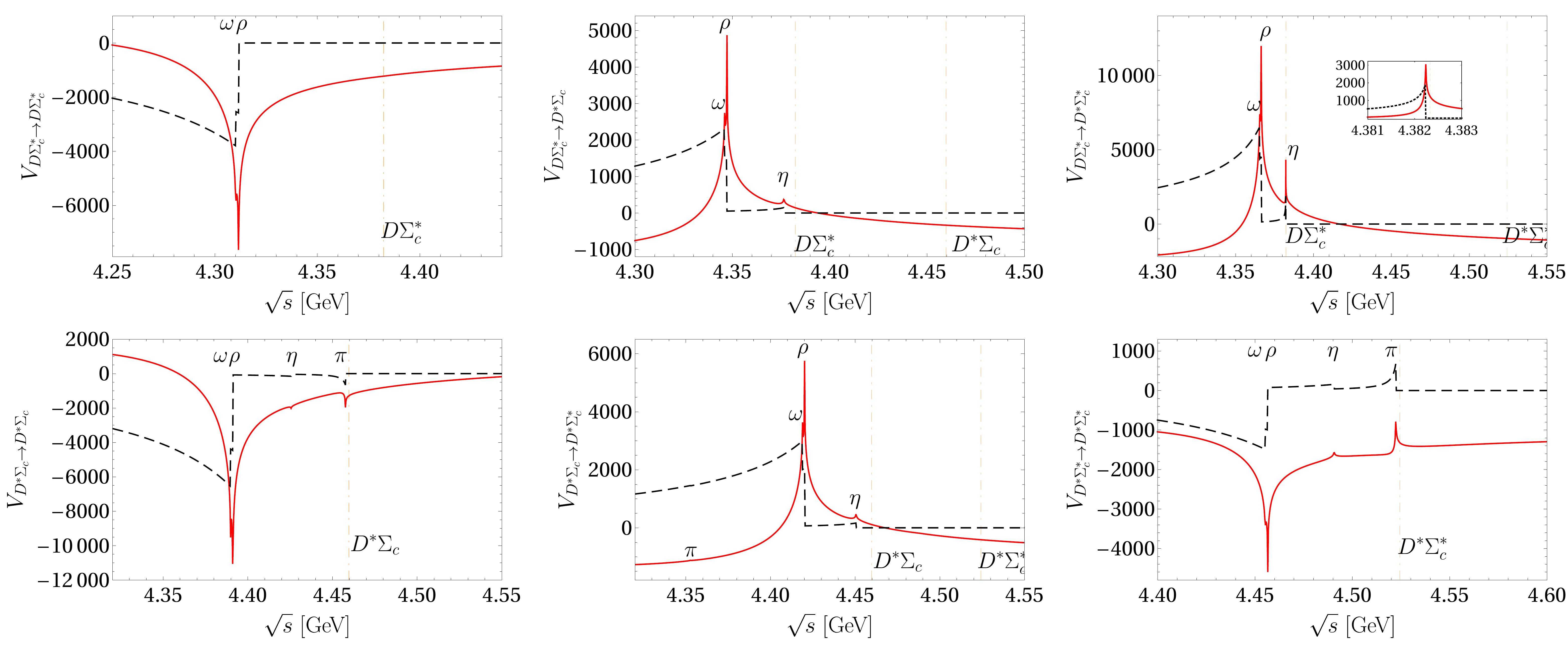}}
		\caption{$V$ matrix for the $S$-wave $D\Sigma_c^*$-$D^*\Sigma_c$-$D^*\Sigma_c^*$
                  coupled-channel system with $(I,J)=(1/2,3/2)$ (real part: solid curve, imaginary
                  part: dashed curve). The corresponding threshold is represented by the orange
                  dash-dotted line.}
		\label{fig:v32}
	\end{figure}

\end{widetext}

\begin{figure}[htbp]
	\centerline{\includegraphics*[width=0.47\textwidth,angle=0]{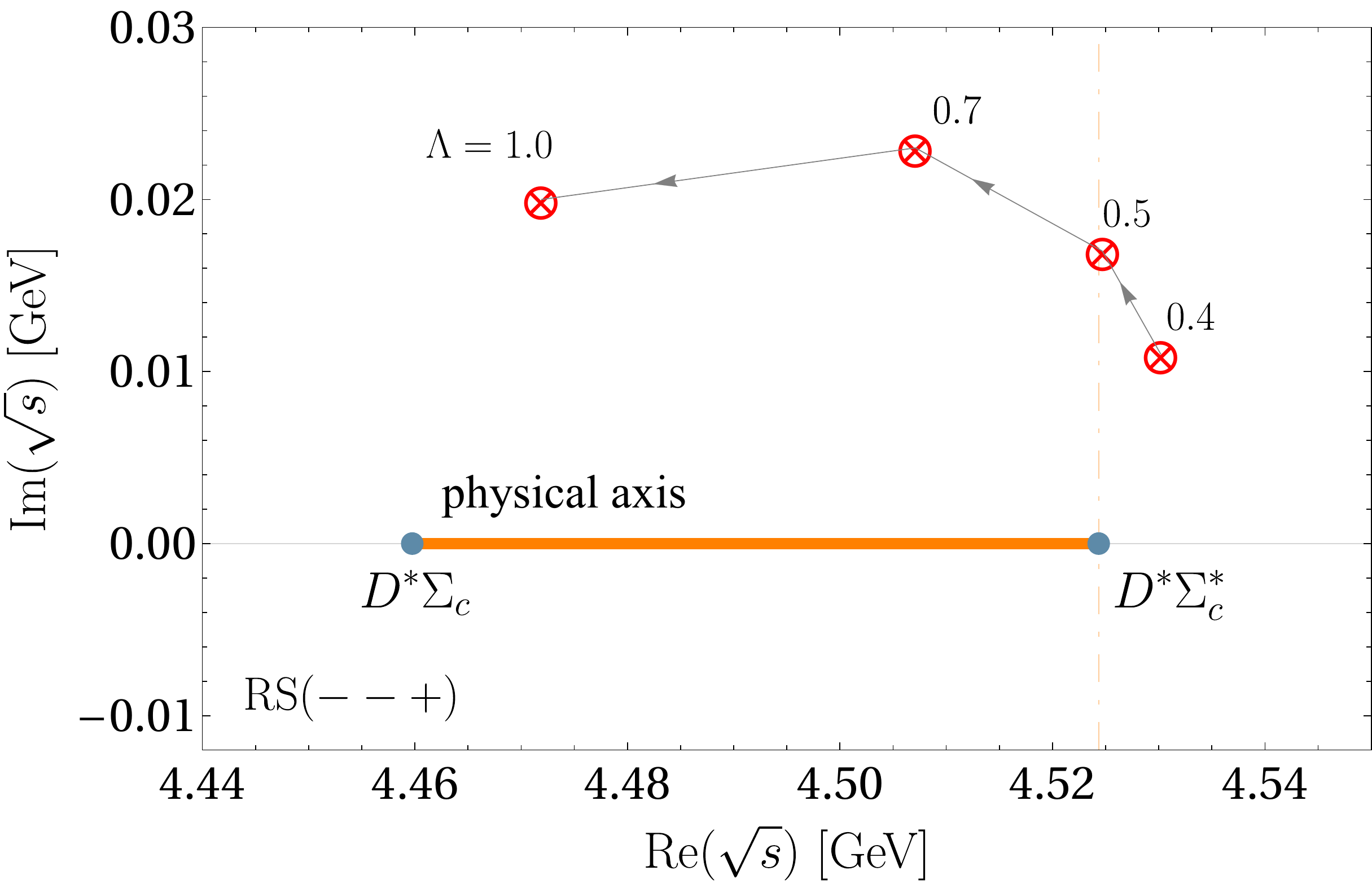}}
	\caption{The pole trajectory on the sheet ${\rm RS}(--+)$ of the $S$-wave
          $D\Sigma_c^*$-$D^*\Sigma_c$-$D^*\Sigma_c^*$ coupled-channel system with $(I,J)=(1/2,3/2)$.
          The cutoff $\Lambda$ is varied from 0.4 to 1.0\,GeV. The highest threshold is represented
          by the orange dash-dotted line. The pole positions are denoted as the red crosses. The
          sheet ${\rm RS}(--+)$ intersects with the physical region at the thick orange line
          that connects the $D^*\Sigma_c$ and $D^*\Sigma_c^*$ thresholds.
	}
	\label{fig:pole32}
\end{figure}

Next, we turn to the $D\Sigma_c^*$-$D^*\Sigma_c$-$D^*\Sigma_c^*$ coupled-channel system with
$(I,J)=(1/2,3/2)$. The $V$ matrix is presented in Fig.~\ref{fig:v32}.
Similar to the $J^P=5/2^-$ single channel case, all the meson-exchange potentials contain the
left-hand cuts after implementing the partial-wave projection except the one-pion-exchange parts
in $V_{D\Sigma_c^*\to D^*\Sigma_c}$ and $V_{D\Sigma_c^*\to D^*\Sigma_c^*}$. 
In particular, the left-hand cuts from the $\eta$-exchange interaction in $V_{D\Sigma_c^*\to D^*\Sigma_c^*}$
and the $\pi$-exchanges in the elastic potentials $V_{D^*\Sigma_c\to D^*\Sigma_c}$ and
$V_{D^*\Sigma_c^*\to D^*\Sigma_c^*}$ are quite close to and below the lower thresholds.
For these cases, the dispersion relations used in $N/D$ method are still valid since no LHC
goes above the RHC.
However, the effects left in the unitarized scattering amplitude $T$ by those near-threshold LHC
are difficult to remove completely.
To do so as much as possible, we shift the energy region, where the subtraction constants in
the $D$ functions of Eq.~\eqref{eq:dfun1} are fitted to functions $(1-VG)$ in Eq.~\eqref{eq:on-shell}
a bit above the corresponding thresholds. In practice, the $D_{ij}$ functions are fitted to
$\delta_{ij}-V_{ij}G_j$ in the energy range starting from $10$\,MeV above the $j$th threshold up
to $100$\,MeV above it.
Then the pole positions are found to be zeros of determinants of $D$ for the physical sheet
and $([T^{\Rom{1}}]^{-1}-2i\rho)$ for the unphysical sheets.
The pole trajectory in the unphysical sheet ${\rm RS}(--+)$ for the
$3/2^-$ $D\Sigma_c^*$-$D^*\Sigma_c$-$D^*\Sigma_c^*$ coupled-channel system is shown in Fig.~\ref{fig:pole32}.
When $\Lambda$ is less than 0.5\,GeV, no $D^*\Sigma_c^*$ bound state with $J^P=3/2^-$ is found.
The cutoff range from 0.6 to 0.8\,GeV is then chosen to give the full mass spectrum in this work.

\begin{widetext}

	\begin{figure}[H]
		\centerline{\includegraphics*[width=0.99\textwidth,angle=0]{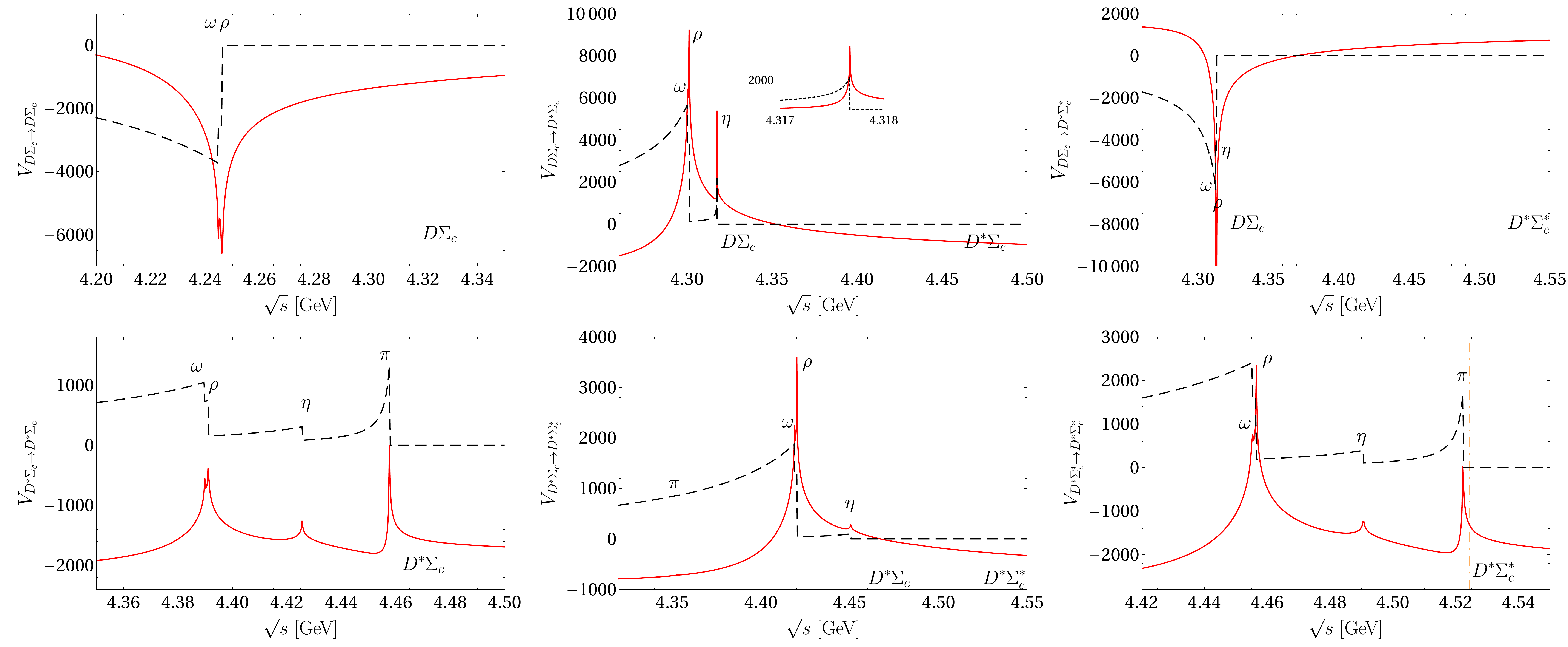}}
		\caption{$V$ matrix for the $S$-wave $D\Sigma_c$-$D^*\Sigma_c$-$D^*\Sigma_c^*$
                  coupled-channel system with $(I,J)=(1/2,1/2)$. For notations, see Fig.~\ref{fig:v32}.
                  }
		\label{fig:v12}
	\end{figure}

\end{widetext}

The calculations for the $D\Sigma_c$-$D^*\Sigma_c$-$D^*\Sigma_c^*$ coupled channels with
$(I,J)=(1/2,1/2)$ can be done along similar lines.
The $V$ matrix for $(I,J)=(1/2,1/2)$ is given in Fig.~\ref{fig:v12}. 

Finally, all the obtained pole positions are collected in the Tab.~\ref{Tab:polepos}.
Similar to other previous works~\cite{Chen:2021htr,Dong:2021bvy,Chen:2021kad}, one $D\Sigma_c$
bound state with $J^P=1/2^-$, one $D\Sigma_c^*$ bound state with $J^P=3/2^-$ and one
$D^*\Sigma_c^*$ bound state with $J^P=5/2^-$ are found.
These three states are located at the physical real axis and below the threshold of the lowest
channel in the corresponding systems, thus they are bound states.
Two resonances located at the complex plane of the unphysical sheets ${\rm RS}(-++)$ and
${\rm RS}(--+)$ are found in both $1/2^-$ and $3/2^-$ coupled channels, which are related
to the $D^*\Sigma_c$ and $D^*\Sigma^*$ bound states with corresponding spin-parities obtained
in the single-channel investigations by Ref.~\cite{Dong:2021bvy}.
The sensitivities of various pole positions to the cutoff are different, the most sensitive
case is the $5/2^-$ $D^*\Sigma_c^*$ pole, whose binding energy varies from $23$\,MeV to $110$\,MeV as
the cutoff changed from $0.6$~\,GeV to $0.8$\,GeV, and the least sensitive one is the $1/2^-$
$D^*\Sigma_c$ pole, whose binding energy varies from $8$\,MeV to $18$\,MeV.
Noted that all the widths of resonances depend on the $\Lambda$ only mildly and thus
the averaged values are listed.

\begin{figure}[htbp]
	\centerline{\includegraphics*[width=0.47\textwidth,angle=0]{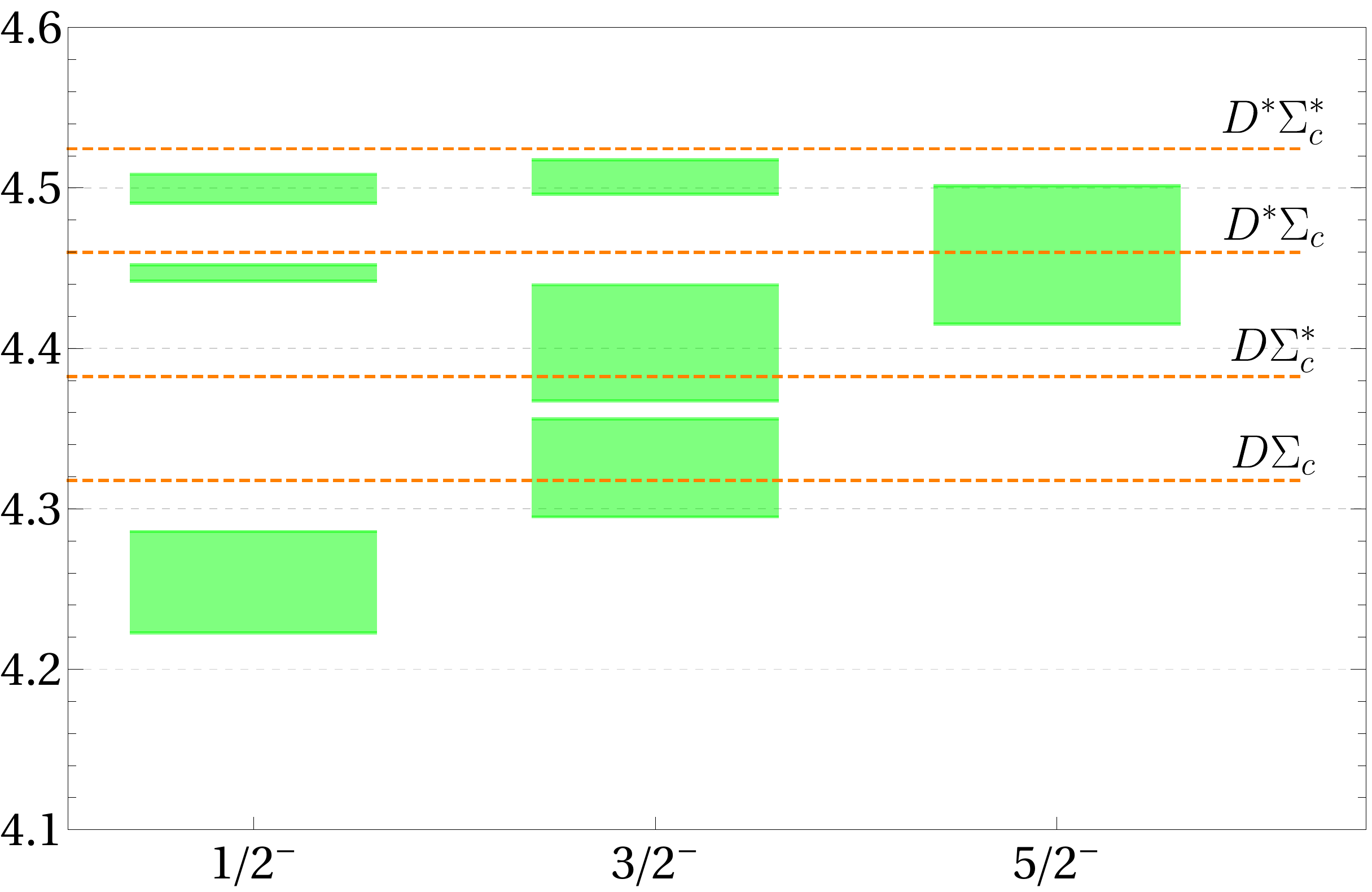}}
	\caption{Mass spectrum of the $S$-wave $D\Sigma_c^*$-$D^*\Sigma_c$-$D^*\Sigma_c^*$
          coupled-channel dynamics with $(I,J)=(1/2,3/2)$, $S$-wave
          $D\Sigma_c $-$D^*\Sigma_c$-$D^*\Sigma_c^*$ system with $(I,J)=(1/2,1/2)$ and
          $S$-wave $D^*\Sigma_c^*$ channel with $(I,J)=(1/2,5/2)$, respectively. The cutoff
          $\Lambda$ is varied from $0.6$\,GeV to $0.8$\,GeV. For the resonances, only the
          real parts of their pole positions are presented.
	}
	\label{fig:spec800}
\end{figure}

We plot the full mass spectrum together with the uncertainty caused by the cutoff $\Lambda$
in Fig.~\ref{fig:spec800}.
The theoretical error from the ambiguity of the matching energy region that needed for the
detemination of the subtraction constants  in Eq.~\eqref{eq:dfun1} is also estimated.
The variation  of this matching range from $(10,100)$\,MeV above threshold to $(30,100)$\,MeV
induces an additional uncertainty of 10\% to 20\% on the binding energies of the various poles.
It should be mentioned that the elimination of those near-threshold left-hand cuts in this work
by employing the first-iterated $N/D$ method is rather qualitative and not a rigorous treatment.
To investigate the mass spectrum in the $D^{(*)}\Sigma_c^{(*)}$ coupled-channel systems more
quantitatively and precisely, a more rigorous and elegant treatment to the uphysical left-hand
cuts is required.
Nevertheless, there is no doubt that the open-charm partners of those LHCb pentaquark states
do exist and are located close to the corresponding $D^{(*)}\Sigma_c^{(*)}$ thresholds.
One can expect those $P_{cc}^N$ states around 4\,GeV will be observed experimentally in the near future.
%

\begin{table}[htbp]
	\centering
	\renewcommand\arraystretch{1.5}
	\caption{Pole positions obtained from the $5/2^-$ $D^*\Sigma_c^*$, $3/2^-$
          $D\Sigma_c^*$-$D^*\Sigma_c$-$D^*\Sigma_c^*$ and $1/2^-$ $D\Sigma_c$-$D^*\Sigma_c$-$D^*\Sigma_c^*$
          systems. The cutoff $\Lambda$ is varied in the range of ($0.6$, $0.8$)\,GeV. All the
          pole positions are given in units of MeV and the corresponding thresholds are listed in the
          brackets in the first column. \label{Tab:polepos}}
	\begin{tabular}{p{1.7cm}<{\centering}|p{2.4cm}<{\centering}|p{2.4cm}<{\centering}|p{1.6cm}<{\centering}}
		\hline
		\hline
		Channel & $1/2^-$ & $3/2^-$ & $5/2^-$ \\
		\hline
		$D\Sigma_c(4318)$  & $(4223, 4286)$  & -	&	- \\
		\hline
		$D\Sigma_c^*(4382)$&	-	&	$(4295,4356)$	&	-	\\
		\hline
		$D^*\Sigma_c(4460)$  & $(4442,4452)+20i$  & $(4367,4440)+1i$	&	- \\
		\hline
		{$D^*\Sigma_c^*(4524)$}  & $(4490,4508)+10i$  & $(4496,4518)+20i$	&	$(4415,4501)$ \\
		\hline
		\hline
	\end{tabular}
\end{table}

\section{Summary} \label{Sec:summary}

Recently, several pentaquark-like states, $P_{\psi}^N$ close to $\bar{D}^{(*)}\Sigma_c^{(*)}$ thresholds
and $P_{\psi s}^\Lambda$ close to the $\bar{D}^{(*)}\Xi_c^{(*)}$, were reported by the LHCb Collaboration.
All these pentaquark states are suggested to be the hadronic molecule candidates composed of the
corresponding mesons close to the pertinent thresholds, see e.g.
Refs.~\cite{Chen:2016qju,Guo:2017jvc,Liu:2019zoy,Brambilla:2019esw,Meng:2022ozq} (and references therein).
The extensions of the mass spectrum of pentaquark states assisted by the heavy quark spin
symmetry and flavor symmetry are expected straightforwardly.
In this work, we explore the possible open-charm partners of $P_{\psi}^N$ states by investigating
the $S$-wave $D^{(*)}\Sigma_c^{(*)}$ coupled-channel dynamics.
Unitarized scattering amplitudes are constructed by means of the first-iterated $N/D$ method.
The open-charm partners of $P_{\psi}^N$ are predicted as follows: for $I(J^P)=1/2(1/2^-)$, one
$D\Sigma_c$ bound state with $E_B=32$\,MeV, one $D^*\Sigma_c$ state with $E_B=8$\,MeV and one
$D^*\Sigma_c^*$ state with $E_B=16$\,MeV; for $I(J^P)=1/2(3/2^-)$, one $D\Sigma_c^*$ bound
state with $E_B=26$\,MeV, one $D^*\Sigma_c$ state with $E_B=20$\,MeV and one $D^*\Sigma_c^*$
state with $E_B=7$~MeV; for $I(J^P)=1/2(5/2^-)$, one $D^*\Sigma_c^*$ state with $E_B=23$~MeV.
Further investigations are required to solidify these results.
%

\section*{Acknowledgments}

We thank  Meng-Lin Du and Yu-Fei Wang for useful discussions and comments.
This work is supported in part by the Deutsche Forschungsgemeinschaft (DFG)
and the National Natural Science Foundation of China (NSFC) through the funds provided to the
Sino-German Collaborative Research Center ``Symmetries and the Emergence of Structure in QCD"
(NSFC Grant No. 12070131001, DFG Project-ID 196253076 -- TRR 110).
The work of UGM was supported in part by The Chinese Academy of Sciences (CAS)
President’s International Fellowship Initiative (PIFI) (grant no. 2018DM0034) and
by the VolkswagenStiftung (Grant No. 93562).

\vspace{1mm}
\begin{appendix}

\section{Effective Lagrangians and exchange potentials} \label{App:formalism}

After expanding the effective Lagrangians in Eq.~(\ref{Eq:Lag}), one has
\begin{widetext}
\begin{align}
    {\cal L}_1 =& \frac{-\sqrt2g}{F_\pi} [ -i \epsilon_{\tau\alpha\mu\nu} v^\alpha P_b^{*(Q)\mu} \partial^\nu \Pi_{ba} P_a^{*(Q)\tau\dagger} + P_b^{*(Q)\mu} \partial_\mu \Pi_{ba} P_a^{(Q)\dagger} + P_b^{(Q)} \partial^\nu \Pi_{ba} P_{a\nu}^{*(Q)\dagger} ], \nonumber \\
    {\cal L}_2 =& \sqrt2 \beta g_V ( P_b^{*(Q)\mu} v^\nu V_{\nu,ba} P_{\mu,a}^{*(Q)\dagger} - P_b^{(Q)} v^\nu V_{\nu,ba} P_{a}^{(Q)\dagger}) - i2\sqrt2 \lambda g_V [ P_b^{*(Q)\mu} ( \partial_\mu V_{\nu,ba} - \partial_\nu V_{\mu,ba} ) P_{a}^{*(Q)\nu\dagger} \nonumber \\
& +i \epsilon_{\tau\lambda\nu\alpha} v^\alpha P_b^{(Q)} \partial^\tau V^\lambda_{ba} P_{a}^{*(Q)\nu\dagger} -i\epsilon_{\alpha\mu\tau\lambda} v^\alpha P_b^{*(Q)\mu} \partial^\tau V^\lambda_{ba} P_{a}^{(Q)\dagger} ], \nonumber \\
    {\cal L}_S =& \frac{-3g_1}{2\sqrt2 F_\pi} v_\kappa \epsilon^{\mu\nu\lambda\kappa} [\bar{B}_{6\mu,ab}^{*(Q)} + \frac1{\sqrt3} \bar{B}_{6,ab}^{(Q)} \gamma_5 \gamma_\mu ] \partial_{\nu} \Pi_{bc} [B_{6\lambda,ca}^{*(Q)} - \frac1{\sqrt3} \gamma_\lambda \gamma_5 B_{6,ca}^{(Q)}] \nonumber \\
& + \frac{\beta_S g_V}{\sqrt2} [\bar{B}_{6\mu,ab}^{*(Q)} + \frac1{\sqrt3} \bar{B}_{6,ab}^{(Q)} \gamma_5 (\gamma_\mu+v_\mu)] v^\alpha V_{\alpha,bc} [B_{6,ca}^{*(Q)\mu} - \frac1{\sqrt3}(\gamma^\mu+v^\mu)\gamma_5 B_{6,ca}^{(Q)}] \nonumber \\
& + \frac{i\lambda_S g_V}{\sqrt2} [\bar{B}_{6\mu,ab}^{*(Q)} + \frac1{\sqrt3} \bar{B}_{6,ab}^{(Q)} \gamma_5 (\gamma_\mu+v_\mu)] (\partial^\mu V^{\nu} - \partial^\nu V^{\mu})_{bc} [B_{6\nu,ca}^{*(Q)} - \frac1{\sqrt3}(\gamma_\nu+v_\nu)\gamma_5 B_{6,ca}^{(Q)}],
\end{align}
in terms of the various field operators 
\begin{align}
    P^{(*)(Q)} &= \sqrt{M} (D^{(*)0}, D^{(*)+}, D_s^{(*)+}), \
    \Pi = \begin{pmatrix} \frac{\pi^0}{\sqrt2}+\frac{\eta}{\sqrt6} & \pi^+ & K^+ \\ \pi^- & -\frac{\pi^0}{\sqrt2}+\frac{\eta}{\sqrt6} & K^0 \\ K^- & \bar{K}^0 & -\sqrt{\frac23}\eta \end{pmatrix}, \
    V = \begin{pmatrix} \frac{\rho^0}{\sqrt2}+\frac{\omega}{\sqrt2} & \rho^+ & K^{*+} \\ \rho^- & -\frac{\rho^0}{\sqrt2}+\frac{\omega}{\sqrt2} & K^{*0} \\ K^{*-} & \bar{K}^{*0} & \phi \end{pmatrix}, \nonumber \\
    B_6^{(Q)} &= \begin{pmatrix} \Sigma_c^{++} & \frac1{\sqrt2}\Sigma_c^{+} & \frac1{\sqrt2}\Xi_c^{\prime +} \\ \frac1{\sqrt2}\Sigma_c^{+} & \Sigma_c^0 & \frac1{\sqrt2}\Xi_c^{\prime 0} \\ \frac1{\sqrt2}\Xi_c^{\prime +} & \frac1{\sqrt2}\Xi_c^{\prime 0} & \Omega_c^0 \end{pmatrix}, \
    B_6^{*(Q)} = \begin{pmatrix} \Sigma_c^{*++} & \frac1{\sqrt2}\Sigma_c^{*+} & \frac1{\sqrt2}\Xi_c^{* +} \\ \frac1{\sqrt2}\Sigma_c^{*+} & \Sigma_c^{*0} & \frac1{\sqrt2}\Xi_c^{* 0} \\ \frac1{\sqrt2}\Xi_c^{* +} & \frac1{\sqrt2}\Xi_c^{* 0} & \Omega_c^{*0} \end{pmatrix}.
\end{align}
\end{widetext}
The values of the coupling constants adopted in the present work are given in Tab.~\ref{Tab:cc}.

\begin{table}[htbp]
    \centering
    \renewcommand\arraystretch{1.4}
    \caption{The coupling constants adopted in the calculation. Here, the values are taken from
      Refs.~\cite{Yalikun:2021bfm,Dong:2021bvy,Chen:2018pzd}. \label{Tab:cc}}
    \begin{tabular}{*{3}{p{1.3cm}<{\centering}}*{2}{p{1.9cm}<{\centering}}}
        \hline
        \hline
        $g_V$ & $g_1$ & $g$ & $\lambda_S$ & $\lambda$ \\
        \hline
        5.8 & 0.94 & -0.59 & -3.31/GeV & 0.56/GeV \\
        \hline
        $\beta$ & $\beta_2$ & $\beta_S$ & $F_\pi$ & \\
        \hline
        0.9 & -0.9 & -1.74 & 0.092\,GeV & \\
        \hline
        \hline
    \end{tabular}
\end{table}

For each process, besides the pseudo-potential ${\cal V}$, which can be derived from the
effective Lagrangians above, there also exists a combined coupling factor (CF) and an
isospin factor (IF).
The combined coupling factor comes from the combination of different charged particles for
the same process.
We collect all the involved combined coupling factors in Tab.~\ref{Tab:if}.
The isospin factors are calculated using~\cite{Wang:2022oof}
\begin{align}
  IF(I) =& \sum_{m_1,m_2,m_3,m_4} \langle I_1 m_1 I_2 m_2 | I I_z \rangle \langle I_3 m_3 I_4 m_4 |
  I I_z \rangle \nonumber \\
  &\times \langle I_3 m_3 I_4 m_4 | M^{iso} | I_1 m_1 I_2 m_2 \rangle,
\end{align}
with the baryon-first convention is applied.
All the isospin factors for the considered processes are also given in Tab.~\ref{Tab:if}.
Then the transition amplitude with specified quantum numbers is expressed by
\begin{equation} \label{Eq:potential}
    V = - (\mathrm{CF}) \times (\mathrm{IF}) \times \mathcal{V}.
\end{equation}
Here the minus sign comes from the definition that ensures a negative $V$ corresponds to
an attraction interaction.

\section{Partial-wave analysis} \label{App:PW}

Once the potential respecting heavy quark spin symmetry is calculated, we can get the
partial-wave potential in the $JLS$ basis using~\cite{Gulmez:2016scm,Wang:2022pin}
\begin{align}
&T^{J}_{L S; L^{\prime} S^{\prime}}(s) = \frac{Y^0_{L^{\prime}}(\hat{\mathbf{z}})}{2J+1}
\sum_{\begin{subarray}{c}   s_{1z}, s_{2z}, s_{3z}, \nonumber \\
s_{4z}, m   \end{subarray}}
\int \mathrm{d} \hat{\mathbf{p}}' Y^m_L (\hat{\mathbf{p}}')^* \nonumber \\
&\phantom{mm}\times (s_{1z} s_{2z} S_z|s_1 s_2 S) (m S_z S_z^{\prime}|L SJ) \nonumber \\
&\phantom{mm}\times (s_{3z} s_{4z} S_z^{\prime}|s_3 s_4 S^{\prime}) (0 S_z^{\prime} S_z^{\prime} | L^{\prime} S^{\prime} J) T(s) \nonumber \\
&\overset{m=0}{=} \frac{\sqrt{(2L^{\prime}+1)(2L+1)}}{2J+1}
\sum_{\begin{subarray}{c}  s_{1z}, s_{2z}, \\ s_{3z}, s_{4z} \end{subarray}}
\frac{1}{2} \int_{-1}^1 \mathrm{d cos \theta} \nonumber \\
&\phantom{mm}\times P_L(\cos \theta) (s_{1z} s_{2z} S_z|s_1 s_2 S) (0 S_z S_z^{\prime}|L SJ) \nonumber \\
&\phantom{mm}\times (s_{3z} s_{4z} S_z^{\prime}|s_3 s_4 S^{\prime}) (0 S_z^{\prime} S_z^{\prime} | L^{\prime} S^{\prime} J) T(s).
\end{align}

Since only the $S$-wave term is considered in the present work, e.g. $L=L^{\prime}=0$, we then have
the following expression
\begin{align}
  T^{J}_{0 S; 0 S^{\prime}}(s) =& \frac1{2J+1}
  \sum_{\begin{subarray}{c}  s_{1z}, s_{2z}, \\ s_{3z}, s_{4z} \end{subarray}}
  \frac{1}{2} \int_{-1}^1 \mathrm{d cos \theta}
  (s_{1z} s_{2z} S_z|s_1 s_2 S) \nonumber \\
  & \times (s_{3z} s_{4z} S_z|s_3 s_4 S^{\prime}) T(s)~.
\end{align}

\begin{table}[h]
	\centering
	\renewcommand\arraystretch{2.0}
	\caption{The combined coupling factors and isospin factors for all the involved processes. It implies the first seven reactions contains the $\pi$, $\eta$, $\rho$ and $\omega$ exchange, while the last two only have the $\rho$ and $\omega$ exchange. \label{Tab:if}}
	\begin{tabular}{p{3.0cm}<{\centering}|p{1.7cm}<{\centering}|p{0.9cm}<{\centering}|*{1}{p{0.9cm}<{\centering}}}
		\hline
		\hline
		Process & \makecell*{Exchanged\\Particle} & CF & IF  \\
		\hline
		\multirow{4}*{\makecell*{$\Sigma_c D \to \Sigma_c D^*$\\
				$\Sigma_c D \to \Sigma_c^* D^*$\\
				$\Sigma_c^* D \to \Sigma_c D^*$\\
				$\Sigma_c^* D \to \Sigma_c^* D^*$\\
				$\Sigma_c D^* \to \Sigma_c D^*$\\
				$\Sigma_c D^* \to \Sigma_c^* D^*$\\
				$\Sigma_c^* D^* \to \Sigma_c^* D^*$}}  & {$\pi$} & {$\frac12$} & {2}  \\
		& {$\eta$} & $\frac16$ & 1  \\
		& {$\rho$} & $\frac12$ & 2  \\
		& {$\omega$} & $\frac12$ & 1  \\
		\hline
		\multirow{2}*{\makecell*{$\Sigma_c D \to \Sigma_c D$\\
				$\Sigma_c^* D \to \Sigma_c^* D$}} & $\omega$  & $\frac12$ & 1  \\
		& $\rho$  & $\frac12$ & 2  \\
		\hline
		\hline
	\end{tabular}
\end{table}
\end{appendix}

\bibliographystyle{plain}

\end{document}